\documentclass[acmsmall,screen,nonacm]{acmart}

\AtBeginDocument{%
  }

\setcopyright{acmlicensed}
\copyrightyear{20XX}
\acmYear{20XX}
\acmDOI{XXXXXXX.XXXXXXX}

\acmConference[XYZ 'XX]{XYZ}{June 03--05,
  20XX}{Woodstock, NY}
\acmISBN{978-1-4503-XXXX-X/18/06}

\usepackage{xspace}
\usepackage{listings}
\usepackage{pgf}
\usepackage{tikz}
\usetikzlibrary{arrows,automata}
\usetikzlibrary{arrows.meta}
\usepackage[vlined,ruled,linesnumbered]{algorithm2e}
\usepackage[noend]{algpseudocode}
\usepackage{caption}
\usepackage{subcaption}
\usepackage{multirow}
\usepackage{hyperref}
\usepackage{pgfplots}
\usepackage{pgfplotstable}
\usepackage{microtype}
\usepackage{balance}
\usepackage{tabularx}
\usepackage{xcolor}
\usepackage{colortbl}
\usepackage{flushend}
\usepackage{enumitem}
\usepackage[norndcorners,customcolors,nofill]{hf-tikz}
\usepackage{xstring}

\newcommand\secref[1]{Sect.~\ref{#1}}

\newcommand\figref[1]{Fig.~\ref{#1}}

\newcommand\tabref[1]{Tab.~\ref{#1}}

\newcommand\circom{\textsc{Circom}\xspace}
\newcommand\noir{\textsc{Noir}\xspace}
\newcommand\corset{\textsc{Corset}\xspace}
\newcommand\gnark{\textsc{Gnark}\xspace}
\newcommand\tool{\textsc{Circuzz}\xspace}
\newcommand\il{\textsc{CircIL}\xspace}
\newcommand{\dollar}{\mbox{\color{BrickRed}\textdollar}}

\newcommand\code[1]{\lstinline[style=basic]{#1}}

\newcommand\bug[1]{
  \IfEqCase{#1}{%
    {A}{\href{https://github.com/iden3/circom/issues/269}{#1}}%
    {1}{\href{https://github.com/iden3/circom/issues/270}{#1}}%
    {2}{\href{https://github.com/iden3/circom/issues/283}{#1}}%
    {3}{\href{https://github.com/iden3/circom/issues/288}{#1}}%
    {4}{\href{https://github.com/iden3/circom/issues/298}{#1}}
    {5}{\href{https://github.com/Consensys/corset/issues/219}{#1}}
    {6}{\href{https://github.com/Consensys/corset/issues/241}{#1}}
    {7}{\href{https://github.com/Consensys/corset/issues/243}{#1}}
    {8}{\href{https://github.com/Consensys/corset/issues/244}{#1}}
    {9}{\href{https://github.com/Consensys/gnark/pull/1181}{#1}}
    {10}{\href{https://github.com/Consensys/gnark/issues/1227}{#1}}
    {11}{\href{https://github.com/Consensys/gnark/pull/1229}{#1}}
    {12}{\href{https://github.com/Consensys/gnark/pull/1234}{#1}}
    {13}{\href{https://github.com/noir-lang/noir/issues/5463}{#1}}
    {14}{\href{https://github.com/AztecProtocol/aztec-packages/issues/8745}{#1}}
    {15}{\href{https://github.com/noir-lang/noir/issues/6150}{#1}}
  }[\PackageError{bug}{Undefined option to bug: #1}{}]%
}

\definecolor{gray}{rgb}{0.5, 0.5, 0.5}
\definecolor{light-gray}{gray}{0.77}
\definecolor{BrickRed}{rgb}{0.8, 0.25, 0.33}
\definecolor{Black}{rgb}{0.0, 0.0, 0.0}
\definecolor{DarkBlue}{rgb}{0.0, 0.0, 0.55}
\definecolor{Crimson}{rgb}{0.86, 0.08, 0.24}
\definecolor{SlateGrey}{rgb}{0.44, 0.5, 0.56}
\definecolor{lightorange}{HTML}{FFB74D}
\definecolor{blue}{rgb}{0.0, 0.0, 1.0}
\definecolor{magenta}{rgb}{0.79, 0.08, 0.48}

\lstdefinestyle{il}{%
  language         = C,%
  alsoletter       = -,%
  morekeywords     = [1]{inputs, outputs, assert},%
  morekeywords     = [2]{},%
  keywordstyle     = [2]\bfseries,%
  morekeywords     = [3]{},%
  keywordstyle     = [3]\bfseries,%
  keywordstyle     = \bfseries,%
  comment          = [l]{//},%
  commentstyle     = \ttfamily\color{Black!60}\footnotesize\lst@ifdisplaystyle\scriptsize\fi,%
  basicstyle       = \ttfamily\footnotesize\lst@ifdisplaystyle\scriptsize\fi,%
  emph             = {},%
  emphstyle        = {\color{teal}\bfseries},%
  stringstyle      = \color{BrickRed},%
  columns          = [c]fixed,%
  aboveskip        = 0mm,%
  belowskip        = 2mm,%
  keepspaces       = true,%
  mathescape       = true,%
  escapechar       = &,%
  tabsize          = 2,%
  numbers          = left,%
  numberstyle      = \tiny\color{Black!70},%
  numbersep        = 4pt,%
  stepnumber       = 1,%
  firstnumber      = 1,%
  showstringspaces = false,%
  captionpos       = b,%
  extendedchars    = true,%
  upquote          = true,%
  abovecaptionskip = 0mm,%
  belowcaptionskip = 0mm,%
  xleftmargin      = 3mm,%
  moredelim        = **[is][{\btHL[fill=light-gray]}]{Â°}{Â°},
  morecomment      = [s]{/*}{*/}
}

\lstdefinestyle{basic}{%
  language         = C,%
  alsoletter       = -,%
  morekeywords     = [1]{signal,input,output,public,template,component,var,function,return,if,else,for,while,do,log,assert,include,parallel,pragma,circom,custom_templates,defcolumns,defconstraint,let,if,pub,fn,new,type,func,nil},%
  morekeywords     = [2]{\~or!,neq!,eq!,is-not-zero!,vanishes!,if-not-zero,is-not-zero,is-zero,to\_be\_bytes,bytes32\_to\_field,SetString,AssertIsEqual,AssertIsLessOrEqual,SetUint64,Neg,Compiler,FieldBitLen},%
  keywordstyle     = [2]\color{teal}\bfseries,%
  morekeywords     = [3]{},%
  keywordstyle     = [3]\color{BrickRed}\bfseries,%
  keywordstyle     = \bfseries\color{DarkBlue},%
  comment          = [l]{//},%
  commentstyle     = \ttfamily\color{Black!60}\footnotesize\lst@ifdisplaystyle\scriptsize\fi,%
  basicstyle       = \ttfamily\footnotesize\lst@ifdisplaystyle\scriptsize\fi,%
  emph             = {},%
  emphstyle        = {\color{teal}\bfseries},%
  stringstyle      = \color{BrickRed},%
  columns          = [c]fixed,%
  aboveskip        = 0mm,%
  belowskip        = 2mm,%
  keepspaces       = true,%
  mathescape       = true,%
  escapechar       = &,%
  tabsize          = 2,%
  numbers          = left,%
  numberstyle      = \tiny\color{Black!70},%
  numbersep        = 4pt,%
  stepnumber       = 1,%
  firstnumber      = 1,%
  showstringspaces = false,%
  captionpos       = b,%
  extendedchars    = true,%
  upquote          = true,%
  abovecaptionskip = 0mm,%
  belowcaptionskip = 0mm,%
  xleftmargin      = 3mm,%
  moredelim        = **[is][{\btHL[fill=light-gray]}]{Â°}{Â°},
  morecomment      = [s]{/*}{*/}
}

\begin{document}

\title{Fuzzing Processing Pipelines for Zero-Knowledge Circuits}

\author{Christoph Hochrainer}
\affiliation{%
  \institution{TU Wien}
  \city{Vienna}
  \country{Austria}}
\email{christoph.hochrainer@tuwien.ac.at}

\author{Anastasia Isychev}
\affiliation{%
  \institution{TU Wien}
  \city{Vienna}
  \country{Austria}}
\email{anastasia.isychev@tuwien.ac.at}

\author{Valentin W\"ustholz}
\affiliation{%
  \institution{ConsenSys}
  \city{Vienna}
  \country{Austria}}
\email{valentin.wustholz@consensys.net}

\author{Maria Christakis}
\affiliation{%
  \institution{TU Wien}
  \city{Vienna}
  \country{Austria}}
\email{maria.christakis@tuwien.ac.at}


\begin{abstract}
Zero-knowledge (ZK) protocols have recently found numerous practical
applications, such as in authentication, online-voting, and blockchain
systems. These protocols are powered by highly complex pipelines that
process deterministic programs, called circuits, written in one of
many domain-specific programming languages, e.g., \circom, \noir, and
others. Logic bugs in circuit-processing pipelines could have
catastrophic consequences and cause significant financial and
reputational damage. As an example, consider that a logic bug in a ZK
pipeline could result in attackers stealing identities or assets. It
is, therefore, critical to develop effective techniques for checking
their correctness.

In this paper, we present the first systematic fuzzing technique for
ZK pipelines, which uses metamorphic test oracles to detect critical logic
bugs. We have implemented our technique in an open-source tool called
\tool. We used \tool to test four significantly different ZK pipelines
and found a total of 16 logic bugs in all pipelines. Due to
their critical nature, 15 of our bugs have already been fixed
by the pipeline developers.
\end{abstract}

\maketitle

\section{Introduction}
\label{sect:intro}

Zero-knowledge (ZK) protocols have recently evolved to enable a wide
range of practical scenarios and
applications~\cite{ErnstbergerChaliasos2024}, including
authentication, online voting, and blockchain systems.
These protocols are called ``zero knowledge'' because they allow one
party, the \emph{prover}, to prove to another party, the
\emph{verifier}, that they know a secret without revealing it.
More specifically, consider a deterministic program $C$, called a
\emph{circuit}, that performs a computation over public and private
(or secret) inputs, $I_P$ and $I_S$ respectively. Given $C$, the
prover must show to the verifier that the computed output $O$ is
indeed produced by executing $C$ with $I_P$ and $I_S$, without however
revealing $I_S$.


\usetikzlibrary {shapes.geometric}


\definecolor{pastel_green}{rgb}{0.800,0.827,0.792}
\definecolor{pastel_blue}{rgb}{0.709,0.752,0.815}
\definecolor{pastel_yellow}{rgb}{0.960,0.909,0.866}
\definecolor{pastel_red}{rgb}{0.933,0.827,0.850}
\definecolor{code_green}{rgb}{0,0.6,0.2}


\tikzset{
  component/.style={
    draw=black!100,
    fill=black!5,
    rounded corners=0.2em,
    align=center,
    line width=0.1em,
    minimum width=7em,
    minimum height=4em,
    font=\large
  },
  artifact/.style={
    fill = pastel_blue,
    draw = black,
    rounded corners=0.2em,
    text = black,
    line width=0.1em,
    minimum width=7em,
    minimum height=3em,
    font=\large
  },
  input component/.style={
    draw=black!100,
    fill=pastel_green,
    rounded corners=0.2em,
    align=center,
    line width=0.1em,
    minimum width=10em,
    minimum height=6em,
    font=\large
  },
  input label/.style={
    font=\large
  },
  input data/.style={
    draw=black!100,
    fill=white,
    align=center,
    shape=circle,
    line width=0.1em,
    minimum width=2.5em,
    minimum height=2.5em,
    font=\large
  },
}

\newcommand{\Cross}{\huge$\mathbin{\tikz [x=1.3ex,y=1.3ex,line width=.2ex, black] \draw (0,0) -- (1,1) (0,1) -- (1,0);}$}%
\newcommand{\Checkmark}{\huge$\color{black}\checkmark$}


\begin{figure}
  \begin{center}
    \resizebox{0.8\linewidth}{!}{
      \begin{tikzpicture}

        \node[draw, component] (compiler)  at (0,0)  {\textbf{Compiler}};
        \draw [-{Triangle[width=8pt,length=3pt]}, line width=3pt](1.25,0) -- (1.75,0);

        \node[draw, component] (generator) at (3,0)  {\textbf{Witness} \\ \textbf{Generator}};
        \draw [-{Triangle[width=8pt,length=3pt]}, line width=3pt](4.25,0) -- (4.75,0);
        \node[draw, artifact]  (witness)   at (6,0)  {\textbf{Witness}};

        \draw [-{Triangle[width=8pt,length=3pt]}, line width=3pt](7.25,0) -- (7.75,0);
        \node[draw, component] (prover)    at (9,0)  {\textbf{Prover}};
        \draw [-{Triangle[width=8pt,length=3pt]}, line width=3pt](10.25,0) -- (10.75,0);
        \node[draw, artifact]  (proof)     at (12,0) {\textbf{Proof}};

        \draw [-{Triangle[width=8pt,length=3pt]}, line width=3pt](13.25,0) -- (13.75,0);
        \node[draw, component] (verifier)  at (15,0) {\textbf{Verifier}};

        \node[draw, input component] (circuit) at (0,2) {};
        \node[input label] (circuit label) at (0,2.5) {Circuit};
        \node[draw, input data] (circuit data) at (0,1.7) {$C$};
        \draw[thick,dash dot,->](circuit data)--(compiler);

        \node[draw, input component] (input) at (3,-2) {};
        \node[input label] (input label) at (3,-2.5) {Inputs};
        \node[draw, input data] (secret input data) at (2.5,-1.7) {$I_S$};
        \draw[thick,dash dot,->](secret input data)--(generator);
        \node[draw, input data] (public input data) at (3.5,-1.7) {$I_P$};
        \draw[thick,dash dot,->](public input data)--(generator);
        \draw[thick,dash dot,->](public input data) to[out=-8,in=-155] (verifier);

        \node (success) at (14.5, -1.5) {\Checkmark};
        \draw[thick,dotted,->](verifier)--(success);
        \node (failure) at (15.5, -1.5) {\Cross};
        \draw[thick,dotted,->](verifier)--(failure);

      \end{tikzpicture}
    } 
  \end{center}
  \caption{Overview of zero-knowledge pipeline stages.}
  \label{fig:overview-zkp}
\end{figure}
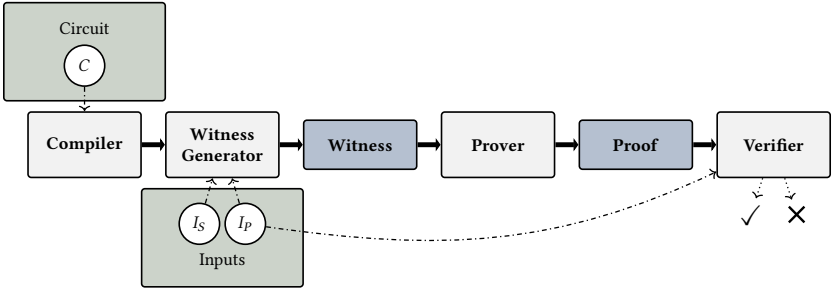

Under the hood, the typical workflow of a ZK pipeline, shown in
\figref{fig:overview-zkp}, is the following. A circuit $C$ is
specified in one of many domain-specific programming languages, such
as \circom~\cite{BellesMunozIsabel2023}, \noir~\cite{Noir},
\corset~\cite{Corset}, and others. The \emph{compiler} of the ZK
pipeline produces a constraint system as well as a \emph{witness
generator}. (Note that, for some pipelines, the witness generator
already exists and is not generated by the compiler.) Given inputs
$I_P$ and $I_S$, the witness generator produces a \emph{witness},
which is essentially an assignment satisfying the constraint system.
On a high level, the witness may also be understood as a trace through
$C$. The witness is used by the \emph{prover} of the pipeline to
generate a proof, which, together with $I_P$, can be passed to the
\emph{verifier} of the pipeline to verify its correctness.

Hence, unlike for regular programs, the pipelines for processing
circuits comprise components for witness and proof generation as well
as for proof verification. They are, therefore, highly complex, and
given their growing applicability, correctness of these pipelines is
highly critical~\cite{ChaliasosErnstberger2024}. More concretely,
consider that the \gnark~\cite{Gnark} and \corset ZK pipelines are
core components of the Linea blockchain, which stores
crypto-assets worth ca. 850M USD as of mid-October 2024.
Bugs in these pipelines could have catastrophic consequences,
potentially causing significant financial and reputational damage.
For this reason, pipeline developers typically follow very strict
development processes. In the case of \gnark~\cite{Gnark}, both
internal and external security teams perform regular audits; 8
external audits alone have been performed in the last two years
(2022--24).

Still, bugs in ZK pipelines are extremely hard to detect just like for
regular compilers and execution environments.
There is, thus, a pressing need to develop automated and effective
techniques for validating their correctness. \emph{Verifying} the
absence of bugs in their implementations is overly
demanding. Consider, for instance, the verification efforts for
CompCert~\cite{Leroy2009}, a compiler for a subset of C---it is about
15K lines of code and required 6 person years to write 100K lines of
specifications. So, for the much more complex ZK pipelines, such an
endeavor is practically infeasible.

\paragraph{Our approach.}
Contrary to verification, automated test-generation techniques have
been previously used to detect bugs in real-world compilers (see
\cite{ChenPatra2020} for an overview) and program analyzers (e.g.,
\cite{ZhangSu2019,MansurWuestholz2023,ZhangPei2023,MordahlZhang2023,HeDi2024,KaindlstorferIsychev2024,FleischmannKaindlstorfer2024})
without providing absolute correctness guarantees, that is, without
promising that all bugs are found.
In this paper, we present \emph{the first fuzzing technique for
finding critical logic bugs in circuit-processing pipelines} and its implementation
in an open-source fuzzer called \tool.

On a high level, \tool generates a configurable number of random
circuits in an intermediate language, which we designed to capture
essential features of many existing ZK languages, such as \circom or
\noir.
Next, \tool applies a sequence of \emph{metamorphic transformations}
to each generated circuit, say $C_1$, to obtain circuit
$C_2$. Metamorphic testing~\cite{ChenCheung1998} typically checks the
correctness of a program $P$ by running the program on an input $i_1$,
observing the output $o_1$, and transforming $i_1$ to obtain
$i_2$. This transformation is such that we know what output $o_2$ to
anticipate (in practice, often the same as for $i_1$), thereby providing an \emph{oracle} for the correct
behavior of $P$. When actually running $P$ on $i_2$, if $o_2$
contradicts the anticipated output, the oracle is violated and a bug
in the program has been found. In our context, inputs $i_1$ and $i_2$
are circuits, the program is a ZK pipeline, and the outputs are the
outputs of the various pipeline stages, e.g., compilation, witness
generation, etc.

After creating transformed circuit $C_2$, e.g., by swapping the
arguments of a commutative operator in $C_1$, \tool translates $C_1$
and $C_2$ from the intermediate language to a specific ZK language. It
then tests the entire processing pipeline, that is, including
compilation, witness generation, proof generation, and proof
verification.
To obtain (public and private) inputs $I_P$ and $I_S$ for the
circuits, \tool generates them randomly. Hence, when executing the
pipeline for each circuit using the same inputs, if any pipeline stage
generates an unexpected output, a bug has been detected. For the
example of swapping the arguments of a commutative operator, we would
not expect any stage outputs to diverge.
This constitutes the first application of metamorphic testing for
circuit-processing pipelines.

In general, the bugs that \tool aims to detect are logic bugs; for
instance, a pipeline stage could be overly permissive (unsoundness), e.g., by
generating a bogus witness or proof, or overly dismissive (incompleteness), e.g., by
not generating a valid witness or proof when it should. As we discuss
in our experimental evaluation, \tool detected 16 logic bugs
in four ZK pipelines, 15 of which are already fixed by the
developers.

Our fuzzing technique primarily works on our intermediate language
instead of specific ZK languages. Note that ZK languages can be quite
different, e.g., \noir is Rust-based whereas \corset is
Lisp-based. The generality that comes with the intermediate language
allows us to easily extend \tool to test new ZK pipelines by adding
the corresponding circuit-generation backends, which translate
circuits from the intermediate to the target ZK language. In fact, we
used \tool to test four significantly different ZK pipelines, namely
\circom~\cite{BellesMunozIsabel2023}, \corset~\cite{Corset},
\gnark~\cite{Gnark}, and \noir~\cite{Noir}.

\paragraph{Contributions.}
Overall, our paper makes the following contributions:
\begin{itemize}
\item We present the first systematic fuzzing technique for
  circuit-processing pipelines; it uses metamorphic test oracles to
  find critical logic bugs.

\item We implement our technique in the open-source tool \tool.

\item We evaluate \tool by testing four different ZK pipelines, namely
  \circom, \corset, \gnark, and \noir; \tool was able to detect logic
  bugs in all pipelines.
\end{itemize}

\paragraph{Outline.}
The rest of the paper is organized as follows. In
\secref{sect:overview}, we give an overview of \tool, and in
\secref{sect:approach}, we describe the technical details of our
technique. \secref{sect:experiments}
presents our experimental evaluation. We review related work in
\secref{sect:related} and conclude in \secref{sect:conclusion}.


\section{Overview}
\label{sect:overview}


\definecolor{pastel_green}{rgb}{0.800,0.827,0.792}
\definecolor{pastel_blue}{rgb}{0.709,0.752,0.815}
\definecolor{pastel_yellow}{rgb}{0.960,0.909,0.866}
\definecolor{pastel_red}{rgb}{0.933,0.827,0.850}
\definecolor{code_green}{rgb}{0,0.6,0.2}


\lstdefinestyle{circom}{
    belowcaptionskip=1\baselineskip,
    breaklines=true,
    frame=none,
    numbers=none,
    basicstyle=\tiny\ttfamily,
    keywordstyle=\bfseries\color{green!40!black},
    commentstyle=\itshape\color{purple!40!black},
    identifierstyle=\color{black},
    backgroundcolor=\color{gray!10!white},
    emph={input,output},
    emphstyle=\bfseries\color{code_green}
}


\tikzset{
  phase/.style={
    draw=black!100,
    fill=black!5,
    rounded corners=0.5em,
    line width=0.1em
  },
  component/.style={
        draw=black!100,
        rounded corners=0.2em,
        align=center,
        line width=0.1em
  },
  generate label/.style={
    fill = white,
    draw = black,
    shape = circle,
    font = \huge,
    minimum size = 0.8cm,
    inner sep = 0.2em,
    line width=0.05em
  },
  circuit/.style={
      sibling distance=4em,
      level distance=2.5em
  },
}


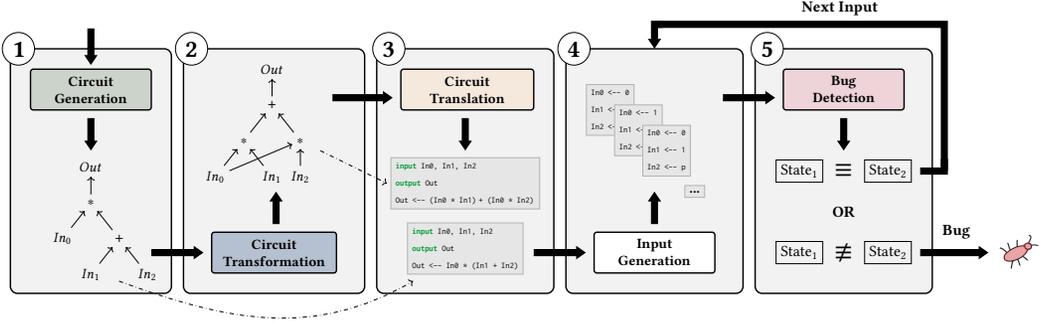
\begin{figure}
  \begin{center}
    \resizebox{\linewidth}{!}{
      \begin{tikzpicture}
        \draw [phase] (-0.5, 1.5) rectangle (3.5, -4.5);
        \draw (-0.3, 1.5) node [generate label] {\textbf{1}};
        \draw [-{Triangle[width=9pt,length=7pt]}, line width=5pt](1.5, 2) -- (1.5, 1.1);
        \draw [component, fill=pastel_green] (0,0) rectangle (3,1)
              node [pos=0.5] (generator) {\textbf{Circuit} \\ \textbf{Generation}};
        \draw [-{Triangle[width=9pt,length=7pt]}, line width=5pt](1.5, -0.2) -- (1.5, -1);
        \node at (1.5, -1.4) {$Out$} [circuit]
          child { node {$*$} edge from parent [<-]
            child { node {$In_0$} edge from parent [<-] }
            child { node {$+$} edge from parent [<-]       
              child { node {$In_1$} edge from parent [<-]}
              child { node {$In_2$} edge from parent [<-]}
            }
          };
        \draw [phase] (3.8, 1.5) rectangle (8.3, -4.5);
        \draw (4, 1.5) node [generate label] {\textbf{2}};
        \draw [-{Triangle[width=9pt,length=7pt]}, line width=5pt](3, -3.5) -- (4.3, -3.5);
        \draw [component, fill=pastel_blue] (4.5, -3) rectangle (7.6, -4)
              node [pos=0.5] (generator) {\textbf{Circuit} \\ \textbf{Transformation}};
        \draw [-{Triangle[width=9pt,length=7pt]}, line width=5pt](6.1, -2.8) -- (6.1, -2);
        \node at (6, 1) {$Out$} [circuit]
          child { node {$+$} edge from parent [<-]
            child { node {$*$} edge from parent [<-]       
              child { node (p2_i0) {$In_0$} edge from parent [<-]}
              child { node {$In_1$} edge from parent [<-]}
            }
            child { node (p2_2nd_times) {$*$} edge from parent [<-]
              child { node {$In_2$} edge from parent [<-]}
            }
          };
        \draw [->] (p2_i0) -- (p2_2nd_times);
      
      \draw [phase] (8.6, 1.5) rectangle (13, -4.5);
      \draw (8.9, 1.5) node [generate label] {\textbf{3}};
      \draw [-{Triangle[width=9pt,length=7pt]}, line width=5pt](7.5, 0.3) -- (9, 0.3);
      \draw [component, fill=pastel_yellow] (9.2, 0) rectangle (12.5, 1)
            node [pos=0.5] (generator) {\textbf{Circuit} \\ \textbf{Translation}};
      \draw [-{Triangle[width=9pt,length=7pt]}, line width=5pt](10.8, -0.2) -- (10.8, -1);
      
      \node [draw=black!30, align=left, fill=gray!20] at (10.8, -1.8) {
          \lstinline[style=circom]{input In0, In1, In2} \\
          \lstinline[style=circom]{output Out} \\
          \lstinline[style=circom]{Out <-- (In0 * In1)} \lstinline[style=circom]{+ (In0 * In2)}};
      \draw [dashdotted, ->](7.2,-1) -- (8.85, -1.8);
      \node [draw=black!30, align=left, fill=gray!20] at (10.8, -3.4) {
          \lstinline[style=circom]{input In0, In1, In2} \\
          \lstinline[style=circom]{output Out} \\
          \lstinline[style=circom]{Out <-- In0 * (In1 + In2)}};
      \draw [dashdotted, ->](2.2,-4.2) to [out=-20,in=210] (9.5, -4.2); 
      \draw [phase] (13.3, 1.5) rectangle (17.7, -4.5);
      \draw (13.5, 1.5) node [generate label] {\textbf{4}};
      \draw [-{Triangle[width=9pt,length=7pt]}, line width=5pt](12.5, -3.5) -- (13.8, -3.5);
      \draw [component, fill=white] (14, -3) rectangle (17, -4)
            node [pos=0.5] (generator) {\textbf{Input} \\ \textbf{Generation}};
      \draw [-{Triangle[width=9pt,length=7pt]}, line width=5pt](15.5, -2.8) -- (15.5, -2);
      \node [draw=black!30, align=left, fill=gray!20] at (14.4, 0) {
          \lstinline[style=circom]{In0 <-- 0} \\
          \lstinline[style=circom]{In1 <-- 1} \\
          \lstinline[style=circom]{In2 <-- p} };
        
      \node [draw=black!30, align=left, fill=gray!20] at (15.1, -0.5) {
        \lstinline[style=circom]{In0 <-- 1} \\
        \lstinline[style=circom]{In1 <-- 1} \\
        \lstinline[style=circom]{In2 <-- p} };
      
      \node [draw=black!30, align=left, fill=gray!20] at (15.8, -1) {
          \lstinline[style=circom]{In0 <-- 0} \\
          \lstinline[style=circom]{In1 <-- 1} \\
          \lstinline[style=circom]{In2 <-- p} };
      \node [draw=black!30, align=left, fill=gray!20] at (16.5, -2) {\textbf{...}};
      \draw [phase] (18, 1.5) rectangle (22.4, -4.5);
      \draw (18.3, 1.5) node [generate label] {\textbf{5}};
      \draw [-{Triangle[width=9pt,length=7pt]}, line width=5pt](17.2, 0.3) -- (18.5, 0.3);
      \draw [component, fill=pastel_red] (18.7, 0) rectangle (21.7, 1)
            node [pos=0.5] (generator) {\textbf{Bug} \\ \textbf{Detection}};
      \draw [-{Triangle[width=9pt,length=7pt]}, line width=5pt](20.2, -0.2) -- (20.2, -1);
      \node at (20.2, -3.5) {\Huge$\not\equiv$};
      \node[draw, align=center] at (21.3, -3.5) {\large State$_2$};
      \node[draw, align=center] at (19.1, -3.5) {\large State$_1$};
      \node at (20.2, -2.5) {\large\textbf{OR}};
      \node at (20.2, -1.5) {\Huge$\equiv$};
      \node[draw, align=center] at (21.3, -1.5) {\large State$_2$};
      \node[draw, align=center] at (19.1, -1.5) {\large State$_1$};
      \draw [line width=5pt](22.1, -1.5) -| (22.7, -1.2);
      \draw [line width=5pt](22.7, -1.5) |- (15.5, 2.1);
      \draw [-{Triangle[width=9pt,length=7pt]}, line width=5pt](15.7, 2.1) -| (15.5, 1.5);
      \node at (20.1, 2.5) {\large\textbf{Next Input}};
      \draw [-{Triangle[width=9pt,length=7pt]}, line width=5pt](22.1, -3.5) -- (23.8,
      -3.5);
      \node at (23.0, -3.0) {\large\textbf{Bug}};
      \node[inner sep=0pt] (bug) at (24.55,-3.5)  {\includegraphics[width=30pt]{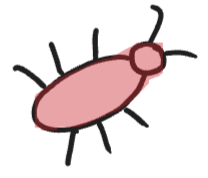}};
      \end{tikzpicture}
    } 
  \end{center}
  \caption{Overview of our fuzzing technique for circuit-processing pipelines.}
  \label{fig:overview}
\end{figure}

\begin{figure}[t!]
\begin{subfigure}[b]{0.5\textwidth}
\begin{lstlisting}[style=il]
inputs : in0, in1 &\label{ir1:in}&
outputs: out0 &\label{ir1:out}&
out0 = (~ &$p$&) &\label{ir1:comp}&
assert(in0 != in1) &\label{ir1:assert}&
\end{lstlisting}
\vspace{-0.5em}
\caption{Circuit $C_1$ in \il.}
\vspace{0.5em}
\label{fig:ex-ir1}
\end{subfigure}%
\begin{subfigure}[b]{0.5\textwidth}
\begin{lstlisting}[style=il]
inputs : in0, in1
outputs: out0
out0 = (~ (((1 - 0) / 1) * &$p$&))
assert(in0 != in1) &\label{ir2:assert}&
\end{lstlisting}
\vspace{-0.5em}
\caption{Circuit $C_2$ in \il.}
\vspace{0.5em}
\label{fig:ex-ir2}
\end{subfigure}
\begin{subfigure}[b]{0.5\textwidth}
\begin{lstlisting}[style=basic]
pragma circom 2.0.6;

template main_template() {
  signal input in0, in1;
  signal output out0;
  out0 <-- (~ &$p$&);
  assert(in0 != in1);
}
component main = main_template();
\end{lstlisting}
\vspace{-0.5em}
\caption{Circuit $C_1$ in \circom.}
\vspace{0.5em}
\label{fig:ex-circom1}
\end{subfigure}%
\begin{subfigure}[b]{0.5\textwidth}
\begin{lstlisting}[style=basic]
pragma circom 2.0.6;

template main_template() {
  signal input in0, in1;
  signal output out0;
  out0 <-- (~ (((1 - 0) / 1) * &$p$&));
  assert(in0 != in1);
}
component main = main_template();
\end{lstlisting}
\caption{Circuit $C_2$ in \circom.}
\vspace{-0.5em}
\label{fig:ex-circom2}
\vspace{0.5em}
\end{subfigure}
\vspace{-1em}
\caption{An example logic bug found by \tool in \circom.}
\label{fig:example}
\end{figure}

In this paper, we propose the first systematic technique for fuzzing
circuit-processing pipelines. An overview of our technique is shown in
\figref{fig:overview}. On a high level, it consists of the following
steps: (1)~circuit generation (in our intermediate language),
(2)~circuit transformation, (3)~circuit translation (to the target
language), (4)~input generation, and (5)~bug detection.
In the rest of this section, we walk the reader through each of these
steps based on an actual logic bug that \tool found in \circom. In
\secref{sect:approach}, we describe each of these steps in detail.

\figref{fig:ex-ir1} shows a circuit, say $C_1$, generated by step~(1)
of our technique. The circuit is expressed in our intermediate
language, which we call \il; lines~\ref{ir1:in}--\ref{ir1:out} declare
the inputs and outputs, line~\ref{ir1:comp} computes output
\code{out0}, and line~\ref{ir1:assert} introduces the constraint that
\code{in0 != in1}.  The output is assigned a constant expression,
namely the bitwise complement of $p$. Here,
$$p =
21888242871839275222246405745257275088548364400416034343698204186575808495617,$$
which is a 254-bit prime number and the base field of the BN254 curve
(also known as alt-BN128), a common prime elliptic curve used in
cryptography. (We omit the value of the prime number in the code due
to its size.)

\figref{fig:ex-ir2} shows circuit $C_2$, again in our intermediate
language, which is generated by step~(2) of our technique, i.e., it is
obtained by applying metamorphic transformations on $C_1$. In
particular, we apply three \emph{equivalence} transformations---that
is, ones that do not alter the semantics of $C_1$---on
line~\ref{ir1:comp}: a multiplication with the identity element
\lstinline[style=basic,mathescape]{(1 * $p$)},
a subtraction of the identity element
\lstinline[style=basic,mathescape]{((1 - 0) * $p$)},
and a division by the identity element
\lstinline[style=basic,mathescape]{(((1 - 0) / 1) * $p$)}.

Step~(3) translates $C_1$ and $C_2$ into the target language, in this
case \circom. The resulting circuits are shown in
\figref{fig:ex-circom1} and \ref{fig:ex-circom2}, respectively. Next,
step~(4) randomly generates input values for signals \code{in0} and
\code{in1}.
Finally, step~(5) tests the entire \circom pipeline on $C_1$ and $C_2$
using the same input values for both circuits. Given that our
metamorphic transformations are semantics preserving, if any
pipeline-stage outputs diverge, a bug is detected.
Specifically, \tool checks whether each stage of the \circom pipeline
(i.e., compilation, witness generation, proof generation, and proof
verification), if executed, succeeds or fails (due to the same
reasons) for both circuits. For the witness-generation stage, \tool
additionally checks whether the generated witnesses are
equivalent. Note that this check is only executed if the generated
input values satisfy the constraint on line~\ref{ir1:assert}.

It is the latter check (of the generated witnesses) that fails when
running \circom on $C_1$ and $C_2$ from \figref{fig:example}. In
particular, \tool found that \code{out0} of $C_1$ evaluates to a very
large number, whereas \code{out0} of $C_2$ evaluates to zero.
Note that the constraint system generated by the compiler from a given
circuit operates over a \emph{finite field}; its size may be
determined by an underlying elliptic curve, and in this case, \tool
randomly selected the BN254 curve. In other words, all arithmetic
operations are computed modulo prime $p$, thereby wrapping around this
value.
The difference in the outputs for $C_1$ and $C_2$ was caused because
\circom did not apply the modulo operation to constants before they
were used in an expression; this can result in computing different
values in the presence of bitwise operations, such as the complement
in our circuits.
Consequently, any one of the above metamorphic transformations alone
would have also revealed the bug.

The developers fixed the issue by applying the modulo operation to all
constants in the abstract syntax tree.


\section{Approach}
\label{sect:approach}

We now describe our fuzzing technique for circuit-processing pipelines
in detail. Before delving into each of its five steps, we provide an
overview of our intermediate language for circuits.

\subsection{Circuit Intermediate Language}
\label{sect:ir}

To easily support fuzzing of diverse circuit-processing pipelines, we
generate and apply metamorphic transformations on circuits expressed
in an intermediate language. This enables reusing the first two steps
of our technique, namely circuit generation and transformation, for
any new pipeline. In other words, to test a new pipeline, it is
primarily the circuit-translation step that must be extended to
translate to the corresponding ZK language. The bug-detection step
also needs to be slightly adapted to execute the new pipeline.

On a high level, our intermediate language, \il, mainly captures a
common subset of many existing ZK languages, such as \circom or
\noir. These languages essentially provide syntactic sugar for
expressing the underlying constraint system in a user-friendly
way. For this reason, the simplest intermediate language could perhaps
be one that directly defines a constraint system, such as R1CS (Rank-1
Constraint System).
However, an intermediate language that is too simple would make it
difficult to test certain features of high-level target languages,
such as \circom, and their corresponding pipelines. We, therefore,
designed a language that is as expressive as possible while still only
supporting features that can easily be translated to many popular ZK
languages.

\il allows defining a circuit using three basic primitives: (1)~a set
of input variables (e.g., line~\ref{ir1:in} of \figref{fig:ex-ir1}),
(2)~a set of output variables (e.g., line~\ref{ir1:out} of
\figref{fig:ex-ir1}), and (3)~a sequence of statements (i.e., the
body), which expresses a set of constraints on the inputs and outputs
(e.g., lines~\ref{ir1:comp}--\ref{ir1:assert} of
\figref{fig:ex-ir1}). The body may contain two types of statements:
(a)~assignments to output variables, where the right-hand side is an
expression over the underlying field (determined by an elliptic
curve), such as line~\ref{ir1:comp} of \figref{fig:ex-ir1}, and
(b)~assertions of Boolean expressions (i.e., the field elements 0 and
1), such as line~\ref{ir1:assert} of \figref{fig:ex-ir1}. Expressions
may be arbitrarily complex by using common operators, such as
addition, multiplication, or equality, supported by all target ZK
languages.

However, some target languages support additional operators. For
instance, the bitwise complement is only used by \circom. As we saw in
\figref{fig:ex-ir1}, \il supports this operator; we provide a list of
all \il operators below.
\begin{description}
\item[Unary operators:] \code{-} (negation), \code{\~} (bitwise complement), and \code{!} (Boolean not)
\item[Binary operators:] \code{+}, \code{-}, \code{*}, \code{/},
  \code{\%} (modulo), \code{**} (power), \code{\&} (bitwise and),
  \code{|} (bitwise or), \code{^} (bitwise xor), \code{\&\&} (Boolean
  and), \code{||} (Boolean or), \code{\^\^} (Boolean xor), \code{==},
  \code{!=}, \code{<}, \code{<=}, \code{>}, and \code{>=}
\item[Ternary operators:] \code{_ ? _ : _} (conditional)
\end{description}
To generate circuits using only operators of the target ZK language,
\tool takes as input a language-specific configuration that enables an
appropriate operator subset.

In general, it is easy to extend our intermediate language to support
language-specific operators. Their translation to the target language
is typically straightforward and enables generating more complex
expressions. After including such operators in \il, we can also use
them to define additional, language-specific metamorphic
transformations.

\subsection{Circuit Generation}
\label{sect:circuit-generation}

The first step of our fuzzing technique randomly generates circuits
based on the intermediate-language grammar.

The circuit-generation component of \tool is highly configurable
allowing users to control the circuit size (by setting the maximum
number of inputs and outputs, the maximum number of assertions, and
the maximum depth of the generated expressions) as well as the circuit
structure (by defining the allowed operators and setting custom
weights determining how often a grammar rule should be applied).
Based on the given configuration, \tool generates a random number of
inputs, outputs, and assertions. It then generates random expressions
to be assigned to outputs and used in assertions. The expressions may
contain constants, inputs, outputs as well as allowed unary, binary,
and ternary operators.

\subsection{Circuit Transformation}
\label{sect:circuit-transformation}

The second step of our technique applies random metamorphic
transformations to a circuit $C_1$ generated by the previous step. The
transformations are designed such that the resulting circuit $C_2$
preserves the semantics of $C_1$.
We can express such transformations using a set of rewrite rules that rewrite a circuit
expressed in the intermediate language.

\paragraph{Rewrite rules.}
We developed a domain-specific language (DSL) for defining rewrite rules
and used it to define a total of 87 rules.

On a high level, the rewrite rules are based on pattern matching. Each
rule is a triple, where the first element provides a unique rule
identifier, the second a pattern to match in the intermediate
language, and the third a rewrite template. For example, the following
rule
\medskip
\begin{lstlisting}[style=basic,numbers=none]
  {"one-plus-zero", "1", "(1 + 0)"}
\end{lstlisting}
is called \code{one-plus-zero} and replaces any occurrence of constant \code{1}
by the expression \code{(1 + 0)}.
We could also generalize the above rule as follows:
\medskip
\begin{lstlisting}[style=basic,numbers=none]
  {"any-plus-zero", "?a", "(?a + 0)"}
\end{lstlisting}

\code{"?<NAME>"} (\code{?a} in the above rule) matches any expression and names it such that it may
be referenced in both the match pattern and rewrite template. For
example, the following rule
\medskip
\begin{lstlisting}[style=basic,numbers=none]
  {"assoc-add", "((?a + ?b) + ?c)", "(?a + (?b + ?c))"}
\end{lstlisting}
employs the associative property of addition and would rewrite an
expression \code{((in0 + 1) + in2)} to \code{(in0 + (1 + in2))}.
Naturally, multiple occurrences of a given name are used to express
structural equality, e.g., the match pattern \code{"?a | ?a"} would
match \code{"(1 + 2) | (1 + 2)"} but not \code{"(1 + 2) | (2 + 1)"}.

Our DSL also allows matching expressions of a given
type. Specifically, \code{"?<NAME>:<TYPE>"} matches any expression of
a particular type, e.g., \code{"?a:bool"} would match \code{1} but not
\code{42}. Recall that all expressions are elements of the field, and
Boolean is a subtype consisting of elements 0 and 1. Therefore, we
currently only need to match type \code{bool}, but in the future, our
DSL could easily be extended to support more
types if necessary. For example, the following rule
\medskip
\begin{lstlisting}[style=basic,numbers=none]
  {"pow2-to-mul", "(?a ** 2)", "(?a * ?a)"}
\end{lstlisting}
rewrites an expression raised to the power of two to the expression
multiplied by itself; no type specification is needed. The following
rule, however,
\medskip
\begin{lstlisting}[style=basic,numbers=none]
  {"double-lor-bool", "?a:bool", "(?a || ?a)"}
\end{lstlisting}
creates a logical disjunction between a Boolean expression and itself.
Note that our intermediate language is untyped, and we perform
Boolean-type inference for pattern matching with such rewrite rules.

Moreover, our DSL allows introducing new random expressions as part of the rewrite template (i.e.,
the third component of the triple). Specifically,
\lstinline[style=basic,mathescape]{"$\dollar$<NAME>:<TYPE>"} is used to generate a random
constant of a given type in the rewrite template and name it. Again, the type specification
is only necessary for generating random Booleans. For example, the following rule
\medskip
\begin{lstlisting}[style=basic,numbers=none,mathescape]
  {"sub-add-random-value", "?a", "((?a - $\dollar$r) + $\dollar$r)"}
\end{lstlisting}
first subtracts a random constant from an expression and then adds it
again; no type specification is needed. The following rule, however,
\medskip
\begin{lstlisting}[style=basic,numbers=none]
  {"double-lxor-bool", "0", "($\dollar$r:bool ^^ $\dollar$r:bool)"}
\end{lstlisting}
replaces constant \code{0} with the exclusive disjunction of a random
Boolean and itself.

In general, using this domain-specific language, we have defined rules
employing the identity, commutative, associative, and distributive
properties of logical, bitwise, and arithmetic operators, De Morgan's
laws, etc.

\paragraph{Stacked transformations.}
To increase the likelihood of finding logic bugs in the tested
pipelines, \tool stacks the above circuit transformations. In other
words, it may apply multiple rewrites to circuit $C_1$ to obtain
$C_2$. This is possible since, in our case, all transformations have the same
equivalence oracle, i.e., that no stage outputs should diverge.

\subsection{Circuit Translation}
\label{sect:circuit-translation}

Once we have two semantically equivalent, but syntactically different
(after applying transformations), circuits $C_1$ and $C_2$, the next
step translates these circuits from the intermediate language to the
ZK language of the processing pipeline under test. \tool currently
supports four diverse ZK languages, namely \circom,
\corset, \gnark, and \noir. In particular, \circom is a low-level
circuit language, which served as an inspiration when designing our intermediate language.
One of its characteristics is that it is modular, thereby allowing users to
define small, parameterizable circuits, called templates (see
\figref{fig:example}), that may then be combined to form larger
circuits. On the other hand, \corset is Lisp-based, \gnark uses plain
Go and provides a high-level API for writing circuits, and \noir is
Rust-based.

This circuit-translation step of \tool is one of the two steps
(besides the bug-detection step) that always needs to be extended when
adding support for a new ZK language. The main task is to map all
supported terminal and non-terminal symbols of the intermediate
language to the corresponding symbols of the ZK language. For instance,
for \gnark, we map \code{a + b} to \code{api.Add(a, b)} and
\code{assert(a <= b)} to \code{api.AssertIsLessOrEqual(a, b)}. For
certain target languages, such as \circom, this step is
straightforward, but for other languages that are quite different from
\il, such as \corset, the translation is more involved.

\subsection{Input Generation}
\label{sect:input-generation}

The input-generation step of \tool produces inputs ($I_P$ and $I_S$)
for circuits $C_1$ and $C_2$. We currently use blackbox fuzzing to
randomly generate elements of the field. However, since the size of
most supported fields is huge, we have introduced the ability to
configure a set of constants that may be used during circuit
generation as interesting boundary values. Examples of such constants
are 0, 1, the size of the field $p$ (typically a prime number), $p -
1$, etc. We randomly pick boundary values with a small probability (e.g., 5\%).
It is due to such a configuration that prime $p$ is used in the circuits
of \figref{fig:example}.

\subsection{Bug Detection}
\label{sect:bug-detection}

As the final step of our technique, bug detection is specific to each
processing pipeline under test and constitutes the other step of \tool
that needs to be extended when adding support for a new ZK
language. For instance, certain pipelines may combine the compilation
and witness-generation stages or support several witness-generation
and proving engines, so their execution requires
configuration. Moreover, detecting bugs in each pipeline involves
retrieving different artifacts, e.g., witnesses, error messages,
return codes, etc., which are non-standard across pipelines.

\paragraph{Oracles.}
The bug-detection step executes the processing pipeline under test for
both circuits $C_1$ and $C_2$ using the same inputs and reports a bug
if our metamorphic oracle is violated, i.e., if any difference is
detected between the two execution behaviors. For instance, a bug is
detected if a witness is obtained only for one of the two semantically
equivalent circuits.
\tool typically executes each pipeline stage for both circuits and
checks for bugs before moving on to the next stage. This allows to
detect any differences in execution behavior as soon as possible
without wasting time on later stages, such as proving or verification,
which can be computationally expensive.

Besides the metamorphic oracles, \tool checks additional
non-metamorphic, correctness properties per execution. For instance, a
bug is reported if the witness generator generates a valid witness,
but the prover fails to produce a valid proof. Similarly, we check
that every valid proof can be successfully verified by the verifier.
In future work, we plan to extend our oracles for the verification
stage, for instance, by also checking if destructive transformations
of generated proofs fail to be verified.

\paragraph{Additional metamorphic transformations.}
Optionally, the bug-detection step may introduce further metamorphic
transformations, not on the input circuits, but on the pipeline
settings. More specifically, we observed that many pipelines support
settings that should not change their functional behavior. For
instance, similar to many C compilers, \circom supports a setting for
selecting the level of optimizations to be applied to the generated
constraint system. Such settings can help in detecting additional bugs
as follows.
We randomly select a pipeline setting $S_1$ for processing $C_1$ and
obtain a setting $S_2$ for processing $C_2$ by applying an equivalent
metamorphic transformation on $S_1$; we expect our oracle to still
hold. For example, in \circom, the \code{--O2} setting, which applies
Gauss elimination to remove as many linear constraints as possible,
could be transformed into the \code{--O0} setting, which disables all
optimizations. These transformations of pipeline settings were
especially effective for \corset, where \tool found 3 logic
bugs due to such transformations.

\subsection{Test-Throughput Optimizations}
\label{sect:discussion}

Certain pipeline stages are computationally expensive, for instance,
proof generation. Consequently, when testing ZK pipelines, the test
throughput is much lower (often seconds or even minutes per test case)
than for many other fuzzing targets, such as parsers (often
milliseconds per test case) or compilers (often less than a few
seconds per test case). However, test throughput is an important
aspect of effective fuzzers. For this reason, \tool implements
optimizations to increase the unusually low throughput when testing ZK
pipelines.

\paragraph{Power schedule.}
An obvious optimization is to skip the slower prover and verifier
stages for some tests, but the interesting question is when to
skip. We initially skipped slower stages randomly with a (fixed) high
probability, but then refined our approach to more fairly test the
initial (faster) stages (i.e., compilation and witness generation) and
the subsequent (slower) stages (i.e., proof generation and
verification). More specifically, \tool is given a target ratio $\rho$
and, for each test, it only executes the later stages if $T_2/(T_1 +
T_2) < \rho$, where $T_1$ is the total time we have already spent
executing the initial stages and $T_2$ is the total time we have
already spent executing the later stages. By default, we use $\rho =
0.5$ to roughly balance the time that is spent in the faster and
slower stages. Over time, \tool dynamically adjusts when to skip the
slower stages, thereby converging toward the desired target ratio.

This approach can be viewed as a novel ``power
schedule''~\cite{BoehmePham2016} assigning energy to different
pipeline stages. The power schedule directly exploits the ZK-pipeline
structure and could be applied to other systems with a similar
structure; for instance, compilers with very expensive optimization
stages, or program verifiers with SMT-based verification stages.

\paragraph{Circuit size and complexity.}
We also observed that the size and complexity of the generated
circuits has a tremendous effect on the performance of the pipeline
stages. In particular, small circuits (e.g., circuits with few
assertions, expressions of small depth, etc.) are preferable since
they tend to be processed much faster. We hypothesize that most bugs
can be detected with small circuits (see RQ3 and RQ4 in
\secref{sect:experiments} for more details). Similar observations have
been made in other domains; for instance, several concurrency testing
tools, such as \textsc{Cuzz}~\cite{BurckhardtKothari2010}, only
explore very few context switches. For this reason, \tool is, by
default, configured to produce smaller and less complex circuits (by
controlling the number of assertions, expression depth, etc.).

Another interesting side effect of large and complex circuits,
especially those with several assertions, is that they make input
generation more challenging---many randomly generated inputs may not
satisfy the assertions. This prevents the fuzzer from effectively
testing later stages of a ZK pipeline, such as proof generation and
verification. For less complex circuits, we can effectively test those
stages with blackbox input generation (see RQ3 and RQ4 in
\secref{sect:experiments} for more details). More complex circuits
would probably require feedback-guided (i.e., greybox) or
whitebox~\cite{GodefroidLevin2008} input generation to satisfy their
constraints. Unfortunately, whitebox fuzzing would likely further
reduce the test throughput, especially due to overhead from constraint
solving. In future work, we plan to explore alternative input
generation techniques and further optimizations.

\paragraph{Circuit bundling.}
Finally, \tool also implements optimizations that are specific to
certain pipelines. For instance, consider that for \gnark, each
circuit needs to be compiled by the regular Go compiler, which is
costlier than the compilation stage of other pipelines. To amortize
this overhead, \tool may bundle multiple circuit pairs (i.e., a
generated and a transformed circuit) that are then (batch-)processed
by the pipeline. In other words, \tool can translate hundreds of
circuit pairs into a single Go test file such that all circuits are
compiled together. Since Go is able to execute individual tests (i.e.,
circuit pairs) in parallel, this bundling optimization has the
additional benefit of parallelizing the bug-detection step in \tool
for free.


\section{Experimental Evaluation}
\label{sect:experiments}

We evaluate \tool by testing four ZK pipelines, namely \circom,
\corset, \gnark, and \noir.  In our evaluation, we address the
following research questions:

\begin{description}
\item[RQ1:] How effective is \tool in detecting logic bugs in diverse
  ZK pipelines?
\item[RQ2:] What are characteristics of the detected bugs?
\item[RQ3:] How efficient is \tool?
\item[RQ4:] How do the design choices and settings of \tool affect its effectiveness?
\end{description}

\subsection{Zero-Knowledge Pipeline Selection}
%
For evaluating the effectiveness and generality of our approach, we
selected four popular, diverse, and maintained ZK pipelines. From a
user perspective, they primarily differ in the ZK language for
specifying circuits, ranging from functional to imperative. However,
they also differ in many technical aspects of the processing stages,
such as the supported constraint systems and cryptographic curves.
Moreover, we chose actively maintained pipelines to ensure that the
developers would respond to any reported bugs. We, therefore, required
the latest activity in their repositories (i.e., commits and responses
to open issues) to be within the last two months.
Next, we provide a high-level overview of each tested pipeline.

\paragraph{\circom.}
At the time of writing, the \circom pipeline has 1.3K stars on GitHub and over 250
forks. It is, for example, used to implement the Tornado cash payment mixer (storing
crypto-assets worth ca. 480M USD as of mid-October 2024). The \circom language is
imperative; it allows operations on constants, input, and output signals, all of which are
field elements.
\circom circuits are compiled to executable
witness generators that, given a set of input signals, can compute
output signals and generate witnesses for the prover and verifier
of the pipeline.

\paragraph{\corset.}
The \corset language is functional and Lisp-like; it provides a
limited set of operations and does not support output signals. Columns
constitute the basic building block of \corset circuits and may be
scalar or array-like; constraints are defined over columns. In
addition to the four common stages, the \corset pipeline has an
optional ``check'' stage that, given field-element assignments to
columns, checks whether the corresponding constraint system is
satisfied.

\paragraph{\gnark.}
At the time of writing, the \gnark pipeline has 1.4K stars on GitHub and over 360
forks. Like \corset, it is currently used to implement the Linea blockchain (storing
crypto-assets worth ca. 850M USD as of mid-October 2024).
Circuits in \gnark can be specified as functions in the (general-purpose) Go
language. Similar to \corset, \gnark does not support output signals;
all signals are considered inputs, which are defined as structs over
field elements.
Moreover, the whole pipeline is embedded in Go, and each stage must be
called using its API.

\paragraph{\noir.}
At the time of writing, the \noir pipeline has over 870 stars on
GitHub and over 180 forks. It provides a strongly-typed, Rust-like
language for specifying circuits. Similar to \circom, \noir supports
explicit output signals that are computed and returned by the
circuits. Unlike the other pipelines, it allows input values other
than field elements. Since this is a unique feature of \noir, we have
not yet added support for it in \tool. Structurally, the \noir
pipeline differs from others by merging the compilation and
witness-generation stages.

\subsection{Experimental Setup}
\label{sect:setup}

\paragraph{Testing time.}
We started testing \circom in March 2024, and we incrementally
improved and extended our fuzzer to support more ZK pipelines. We
subsequently added support for \gnark (in June 2024), \corset (in July
2024), and \noir (also in July 2024).
As shown by this timeline, once \tool was mature enough, we were able
to add new pipelines without too much effort.

Due to this timeline however, we did not spend the same amount of
total fuzzing time on each pipeline.
We estimate that we fuzzed \circom for ${\sim}5$ months, \gnark for
${\sim}4$ months, and \corset and \noir for ${\sim}3$ months. Note
that, once a bug was detected, we typically did not continue fuzzing
the corresponding pipeline until the bug was fixed to avoid reporting
duplicate issues.

For all pipelines, we tested either the latest stable release or the
main development branch (to potentially find bugs that were introduced
more recently).

\paragraph{\tool settings.}
Over time, we refined the default setup for \tool based on our
experience. In particular, we identified the following key settings
and default values: (1)~the maximum number of inputs and outputs (each
defaulting to 2), (2)~the maximum number of assertions (defaulting to
2), (3)~the maximum expression depth (defaulting to 4), and (4)~the
maximum number of stacked transformations that are applied to a
generated circuit (defaulting to 64). The former three may affect the
size and complexity of the circuits, and thus, the test
throughput. The latter aims to strike a balance between finding bugs
faster (by applying more transformations) and facilitating debugging
(by not producing transformed circuits that differ too much from the
generated ones).
As discussed in \secref{sect:discussion}, we use a default target
ratio of $\rho = 0.5$ to roughly balance the time that is spent in the
initial (typically faster) stages (i.e., compilation and witness
generation) and the subsequent (typically slower) stages (i.e., proof
generation and verification).

In RQ4, we compare different configurations of these settings in terms
of their bug-finding effectiveness. To this end, we evaluate which
configurations are able to refind bugs that we reported to the
pipeline developers. To ensure that a bug detected by a given
configuration indeed corresponds to the original, reported bug (and
not to another one), we apply the fix that was provided by the
developers and check whether the buggy behavior
disappears. For this reason, we only use fixed bugs for evaluating the
effectiveness of different \tool configurations.

\paragraph{Fuzzing campaigns.}
To ensure a fair comparison and mitigate the effects of randomness in
the fuzzing process, we run 10 independent fuzzing campaigns for each
\tool configuration. We limit the duration of each campaign to 24
hours. In general, we did not limit the time per pipeline
execution. However, even though \corset was generally one of the
fastest pipelines, we observed that it would occasionally (i.e., only
for a few circuits) use over 1TB of memory and take several hours to
run. For this reason, we introduced an upper bound of 8GB memory usage
per pipeline execution (only for \corset).

\paragraph{Hardware.}
We performed all experiments on a machine with an AMD EPYC 9474F CPU @
3.60GHz and 1.5TB of memory, running Debian GNU/Linux 12
(bookworm). To avoid issues due to hardware resources and obtain
reproducible results, we restricted each fuzzing campaign to use a
single logical CPU core.

\subsection{Experimental Results}

We now discuss our findings for each research question.

\paragraph{RQ1: Effectiveness of \tool.}
%
\tabref{tab:bugs} shows all unique bugs found by \tool in the ZK
pipelines we tested. The first column assigns an identifier (ID) to
each bug and links to the bug report. We assign a number to fixed bugs
and a letter to others. The second and third columns show the ZK
pipeline in which the bug was found, and the bug status (i.e.,
reported, confirmed, or fixed). The fourth and fifth columns provide
the pipeline stage where the bug was detected, and the oracle that
detected it. Here, ``MT'' denotes a metamorphic oracle, and ``VC''
stands for validity check, i.e., a non-metamorphic, correctness
property asserting the successful execution of a pipeline stage (see
\secref{sect:bug-detection}). The last column includes a short
description of the bug.

\emph{In total, \tool detected 16 unique bugs, 15 of which were
previously unknown.} Bug~\bug{13} was found in the latest \noir
release, but the developers had independently detected and fixed it in
their development branch.
Recall from \secref{sect:intro} that ZK pipelines are regularly
audited, and their developers follow strict procedures; yet, \tool was
effective in detecting previously unknown, logic bugs. \emph{13 of the bugs
were detected due to violating a metamorphic oracle and 3 due to
violating a validity check.}
Of the 13 that were detected due to violating a metamorphic oracle, 10
involved metamorphic transformations on the circuits and 3 on the
pipeline settings (see \secref{sect:approach}). The latter
transformations (on the pipeline settings) uncovered
bugs~\bug{5},~\bug{6}, and~\bug{7} in \corset.
\tool also found several crashes as a by-product, but we did not
report most of them to focus on critical issues. For instance, when
reporting bug~\bug{5}, we discovered another, less severe bug, and
developers opened an independent
\href{https://github.com/Consensys/corset/issues/227}{issue} to track
it.

\begin{table}[t]
    \small
    \caption{Unique logic bugs detected by \tool.}
    \resizebox{\textwidth}{!}{%
    \begin{tabular}{c|c|c|c|c|l}
    \begin{tabular}[c]{@{}c@{}}\textbf{Bug}\\\textbf{ID}\end{tabular} & \textbf{Pipeline}
    & \textbf{Status} & \textbf{Stage} & \textbf{Oracle} &\textbf{Description} \\
    \hline
    \bug{A} & \circom & confirmed &  {Prover} & VC &``Polynomial is not divisible'' error \\
    \bug{1} & \circom & fixed & {Witness} & MT & Incorrect evaluation of bitwise complement of constants \\
    \bug{2} & \circom & fixed & {Witness} & MT & Incorrect evaluation of bitwise complement of zero \\
    \bug{3} & \circom & fixed & {Witness} & MT & Inconsistent evaluation of field prime \\ 
    \bug{4} & \circom & fixed & {Witness} & MT & Inconsistent evaluation of a small field prime \\ 
    \hline
    \bug{5} & \corset & fixed & {Check} & MT & Inconsistent behavior of expansion and native flags \\
    \bug{6} & \corset & fixed & {Check} & MT & Incorrect evaluation of constraints using expansion \\
    \bug{7} & \corset & fixed & {Check} & MT & Incorrect expansion transformation of conditionals \\ 
    \bug{8} & \corset & fixed & {Check} & MT & Incorrect evaluation of normalized loobean \\
    \hline
    \bug{9} & \gnark & fixed & {Witness} & MT & Inconsistent evaluation of $\lor$ for constants and signals \\
    \bug{10} & \gnark & fixed & {Witness} & MT & Incorrect evaluation of {\footnotesize\texttt{AssertIsLessOrEqual}} \\
    \bug{11} & \gnark & fixed & {Compiler} & MT & Zero bit length for binary decomposition on constants \\
    \bug{12} & \gnark & fixed & {Compiler} & MT & Compiler panic on branch with unchecked cast \\ 
    \hline
    \bug{13} & \noir & fixed & {Witness} & MT & Incorrect evaluation of asserted condition \\
    \bug{14} & \noir & fixed & Prover & VC & Proof failure due to insufficiently large string \\
    \bug{15} & \noir & fixed & Compiler & VC & Stack overflow for $<$ with nested expressions \\
    \end{tabular}
    }
\label{tab:bugs}
\end{table}

\emph{15 bugs are already fixed by the pipeline developers, attesting
to their critical nature}.  Addressing bug~\bug{A} is ``\emph{quite a
challenge}'' for the developers, which is why it has not yet been
fixed.
Even though most of our bugs were fixed quickly, some of them within
hours of our report, they were often non-trivial to address. For
instance, for bugs~\bug{5} and~\bug{10}, the initial proposed fixes
addressed the underlying problem only partly, and \tool quickly
uncovered follow-up issues~\bug{6} and~\bug{11}, respectively, which
required additional changes in the code.

As shown in the table, 3 bugs were found in the compilation stage, 7
in the witness-generation stage, and 2 in the proof-generation stage.
\tool found all 4 \corset bugs when executing the optional check
stage, which checks the validity of the constraint system generated
from a given circuit. Note that we run the \corset check stage before
the compilation stage. We find it encouraging that \tool found most
issues in early pipeline stages, which is likely due to the fact that
the proof-generation and verification stages are audited even more
thoroughly. Additionally, since these stages are significantly more
computationally expensive than others, we ran them less frequently
(according to our target ratio $\rho$). While we can configure \tool
to execute the full pipeline more often, that would significantly
decrease test throughput (see RQ4).

The feedback from the pipeline developers was overwhelmingly
positive. For instance, one of the main developers of \gnark responded
with ``\emph{that fuzzer is killing it!}'' when we reported
bug~\bug{12}. In response to bug~\bug{8}, \corset developers called it
a ``\emph{critical bug actually. Good spotting!}''. It turned out that
a feature to support word-wise normalization was only partially
implemented, but it was used in the standard library. Overall, all
teams strongly encouraged us to keep fuzzing their code.

\paragraph{RQ2: Detected logic bugs.}
%
In the following, we provide a more detailed description of bugs found by \tool in each of
the tested pipelines. Note that, for simplicity, we show manually minimized versions of
the generated and transformed circuits. In practice, we also manually minimized the
circuits that we included in our bug reports. This tends to make it much easier for
developers to debug and fix the issues.

\begin{figure*}[t!]
\begin{subfigure}[b]{.49\textwidth}
\begin{lstlisting}[style=basic]
template C1() {
  signal a;
  a <-- (~ 0);
}

template C2() {
  signal a, tmp, zero;
  tmp <-- (0 ^ 0); &\label{code:circom:tmp1}&
  zero <-- (~ tmp); &\label{code:circom:tmp2}&
  a <== zero;
}
\end{lstlisting}
\caption{Bug~\bug{2} in \circom.}
\label{fig:circom-bug}
\end{subfigure}
\begin{subfigure}[b]{.49\textwidth}
\begin{lstlisting}[style=basic,alsoletter=!-~]
(defcolumns in0)
(defconstraint C1 ()
  (vanishes! (let ((out0 in0))
    (let ((out1 0)) &\label{code:corset:dup1}&
      (eq!
        (is-not-zero!  &\label{code:corset:dup2-start}&
          (eq!  &\label{code:corset:dup3-start}&
            (if (is-zero out1) out1 1) &\label{code:corset:dup3-end}&
            (neq! in0 in0))) &\label{code:corset:dup2-end}&
        (~or! 0 out0))))))
\end{lstlisting}
\caption{Bug~\bug{6} in \corset.}
\label{fig:corset-bug}
\end{subfigure}
\caption{Critical bugs detected by \tool in \circom (left) and \corset (right) and fixed by the developers.}
\label{fig:bugs1}
\end{figure*}

\figref{fig:circom-bug} shows two circuits that revealed
bug~\bug{2}. \tool generated circuit \code{C1} (top) that computes the
bitwise complement of 0. It then transformed \code{C1} into the
equivalent circuit \code{C2} (bottom) by replacing \code{0} with
\code{0 ^ 0}. \circom only allows quadratic constraints, which
\lstinline[style=basic]{~ (0 ^ 0)} is not, therefore \tool
(internally) rewrites this expression into the intermediate
assignments on lines~\ref{code:circom:tmp1} and
\ref{code:circom:tmp2}. When executed, the two circuits computed
different values for signal \code{a}.
The discrepancy came from a difference in the sign of 0 and
\code{zero}: constant 0 was considered positive, while signal
\code{zero} negative, resulting in different bitwise
complements. The developers fixed the issue by enforcing a positive sign
on all zero operands of the bitwise complement. Interestingly,
bug~\bug{3} presented earlier (see \figref{fig:example}) was
discovered by a syntactically similar pair of circuits but uncovered a
distinct issue, as we explain in \secref{sect:overview}.

\figref{fig:corset-bug} shows bug~\bug{6} found in \corset; there is
no need to understand the functionality of the code other than observe
that there is an if-condition nested within an expression on
line~\ref{code:corset:dup3-end}. This issue was discovered by running
\corset on a single circuit, namely \code{C1} from \figref{fig:corset-bug}, but
with different flags. \tool first ran the pipeline with the \code{-N}
flag, which enables native mode. By default, all operations are
performed using \code{BigInt} objects, and native mode uses
(mathematical) field values instead. \tool then transformed this
setting into the \code{-Ne} flag. The \code{e} part of the flag
enables expansion mode, which rewrites constraint expressions into a
lower-level, but equivalent, form. Even though flag \code{-e} should not
affect the constraint satisfiability, the constraints were found SAT in
native mode but UNSAT when enabling both the native and expansion
modes.

After closer inspection, the developers responded that ``\emph{there
is a problem with the handling of if-conditions when they are nested
within certain expressions}''. The problematic part of the code, which was
only triggered when enabling expansion mode, intended to hoist
the nested if-conditions into separate constraints. The proposed fix
changed the order of cases when pattern matching a nested
if-condition. However, when testing the fixed version, \tool revealed
that, even though the fix worked for the provided circuit, there were
still cases where \corset did not treat if-conditions correctly.
Based on this finding, we reported bug~\bug{7}. The developers
concluded that ``\emph{the related issue is in the same (source-code)
file as the original problem, but in a different method}''. To fix the
issue, they had to rework the handling of nested if-conditions (in
expansion mode) from scratch.

\begin{figure*}
\begin{subfigure}[b]{.49\textwidth}
\begin{lstlisting}[style=basic]
func (circuit *C1) Define(api frontend.API
                            ) error {
  api.AssertIsLessOrEqual(1, 0) &\label{code:gnark-replace1}&
  return nil
}

func (circuit *C2) Define(api frontend.API
                            ) error {
  api.AssertIsLessOrEqual(1, api.Or(0, 0)) &\label{code:gnark-replace2}&
  return nil
}
\end{lstlisting}
\caption{Bug~\bug{10} in \gnark.}
\label{fig:gnark-bug}
\end{subfigure}
\begin{subfigure}[b]{0.49\textwidth}
\begin{lstlisting}[xleftmargin=\parindent,style=basic,deletekeywords={input}]
fn main(input : Field) -> pub Field {
  let b2 : [u8; 32] = input.to_be_bytes();
  let b2_f =
    std::field::bytes32_to_field(b2); 
  assert(0 != b2_f, "Assertion violated");
  0
}\end{lstlisting}
\caption{Bug~\bug{14} in \noir.}
\label{fig:bug-noir-prover}
\end{subfigure}
\caption{Critical bugs detected by \tool in \gnark (left) and \noir (right) and fixed by the developers.}
\Description{}
\label{fig:bugs2}
\end{figure*}

\tool discovered an issue in the evaluation of the
\code{AssertIsLessOrEqual} primitive in \gnark that we reported as
bug~\bug{10}. Consider circuit \code{C1} shown in
\figref{fig:gnark-bug}. For \code{C1}, \gnark failed to generate a
witness since the assertion does not hold. Then, \tool applied the
following rewrite rule
\smallskip
\begin{lstlisting}[style=basic,numbers=none]
  {"zero-or", "?a", "(?a | 0)"}
\end{lstlisting}
that transformed \code{0} (line~\ref{code:gnark-replace1}) into the
equivalent expression \code{api.Or(0, 0)}
(line~\ref{code:gnark-replace2}). Unlike for \code{C1}, \gnark
successfully generated a witness for the transformed circuit
\code{C2}. The \gnark developers modified the assertion API code to
correctly handle the special case where the first argument of
\code{AssertIsLessOrEqual} is a constant.

This fix, however, overlooked another corner case where the first
constant in the assertion is zero. For instance, when replacing
\code{1} by \code{0}, the proposed fix did not work, and the
metamorphic oracle still failed. \tool found this corner case in a
subsequent fuzzing campaign (bug~\bug{11}). The final fix removed the
code that was trying to optimize the evaluation of such assertions for
constant arguments. The two iterations that were required to fix the
root issue suggest that, despite thorough testing on the developer
side (including unit and regression tests), no test was able to catch
neither the initial issue (bug~\bug{10}), nor the issue that remained
after the partial fix (bug~\bug{11}). This provides a glimpse into the
complexity that developers of ZK pipelines are facing and highlights
the need for automated test generation.

\figref{fig:bug-noir-prover} shows the circuit that revealed
bug~\bug{14} in the \noir prover. Before generating a proof, \noir
creates a structured reference string (SRS), which records proving and
verification parameters as well as a sequence of samples from some
complex (secret) distribution. This string is later used to verify the
correctness of the proof. The number of required samples in the SRS
depends on the circuit and is estimated automatically. When proving
our example circuit, the \noir pipeline crashed because the
(automatically) estimated number of samples in the SRS was too
small. This bug was revealed as a violation of our validity check that
expects a successful witness generation to be followed by a successful
proof. Developers fixed the issue by changing the algorithm to
over-approximate the number of necessary samples.

Overall, the presented bugs demonstrate that \tool is able to identify
a diverse---for instance, with respect to the different ZK pipelines,
oracles, and affected pipeline stages---set of critical bugs that
developers are eager to fix.

\paragraph{RQ3: Efficiency of \tool}
%
We primarily evaluate the efficiency of \tool in terms of its
\emph{bug-finding time}. We additionally measure the number of
circuits that had to be generated to find a given bug. We track these
metrics for each \emph{fixed} issue discovered by \tool (listed in
\tabref{tab:bugs}). We only use fixed issues for this evaluation since
the difference in behavior of buggy and fixed code provides a reliable
way to identify if a given bug was indeed detected by the fuzzer.

\begin{table}[t]
\caption{Time and number of generated circuits that the default
  configuration of \tool needed to find a fixed issue across 10
  independent fuzzing campaigns, each with a time limit of 24 hours.}
  \scalebox{0.93}{
  \begin{tabular}{c|c|c|c|rrr|rrr}
  \textbf{ZK} & \textbf{Bug} & \multirow{2}{*}{\textbf{Seeds}} & \textbf{SAT} & \multicolumn{3}{c|}{\textbf{Time to bug}} & \multicolumn{3}{c}{\textbf{Circuits to bug}} \\
  \textbf{Pipeline} & \textbf{ID} & & \textbf{inputs} & \multicolumn{1}{c}{\textbf{min}} &
  \multicolumn{1}{c}{\textbf{med}} & \multicolumn{1}{c|}{\textbf{max}} & \multicolumn{1}{c}{\textbf{min}}
  & \multicolumn{1}{c}{\textbf{med}} & \multicolumn{1}{c}{\textbf{max}} \\
  \hline
\multirow{4}{*}{\circom}
& \bug{1}      & 10    & 52.88\%   & 38s      & 4m14s       & 13m47s    & 7        & 61        & 212 \\
& \bug{2}      & 10    & 57.84\%   & 13m08s    & 31m00s         & 1h00m30s     & 232      & 567       & 1265 \\
& \bug{3}      & 10    & 57.38\%   & 43s      & 13m30s      & 35m17s    & 19       & 210       & 717 \\ 
& \bug{4}      & 10    & 57.53\%   & 18s      & 7m19s       & 16m13s    & 8        & 108      & 247 \\
\hline
\multirow{4}{*}{\corset}
& \bug{5}      & 10    & 65.08\%   & 1s       & 17s         & 2m24s     & 3        & 11        & 66 \\ 
& \bug{6}      & 10    & 64.21\%   & <1s      & 20s         & 2m45s     & 5        & 17        & 117\\ 
& \bug{7}      & 10    & 61.17\%   & 1s       & 3m35s       & 12m11s    & 5        & 91       & 340 \\
& \bug{8}      & 10    & 63.26\%   & 30s      & 3m26s       & 11m56s    & 19       & 112       & 377 \\
\hline
\multirow{4}{*}{\gnark} 
& \bug{9}      & 10    & 57.05\%   & 1m54s    & 11m28s      & 25m12s    & 10       & 122       & 372\\
& \bug{10}     & 10    & 55.97\%   & 1m13s    & 13m13s      & 44m21s    & 9        & 141       & 663\\
& \bug{11}     & 10    & 56.94\%   & 2m01s     & 13m57s      & 39m17s    & 21       & 162       & 634\\
& \bug{12}     & 10    & 55.68\%   & 30m20s   & 2h26m54s    & 20h21m27s & 446      & 2361      & 15329\\
\hline
\multirow{3}{*}{\noir}  
& \bug{13}     & 10    & 59.31\%   & 29m45s   & 57m12s      & 5h07m37s   & 399      & 843       & 4963\\
& \bug{14}     & 10    & 58.88\%   & 1h55m37s & 10h17m13s   & 16h38m51s & 488      & 2908      & 4780\\
& \bug{15}     & 10    & 59.05\%   & 8m19s    & 48m40s      & 1h49m52s  & 25       & 184       & 450\\
  \end{tabular}}
\label{tab:efficiency}
\end{table}

\tabref{tab:efficiency} summarizes the results of this experiment
across 10 independent fuzzing campaigns, i.e., using 10 different
random seeds to randomize the fuzzing process, each with a time limit
of 24 hours.
We use the default configuration of \tool as described in
\secref{sect:setup}, i.e., up to 2 inputs, outputs, and assertions per
circuit, expressions of depth up to 4, and up to 64 stacked
transformations.
The first two columns of the table show the ZK pipeline under test and
the unique bug IDs (from \tabref{tab:bugs}), and the third column
indicates for how many of the independent fuzzing campaigns (i.e.,
random seeds) \tool found the given bug.  The fourth column presents
the percentage of generated circuit inputs that satisfy the
corresponding constraint system.
The remaining columns show the minimum, median, and maximum bug-finding
times across all campaigns as well as the minimum, median, and maximum
number of circuits that were generated until each bug was found.

\emph{\tool reliably detects all issues in all 10 campaigns, and the
median bug-finding time for 13 (out of 15) bugs is less than 1
hour. The median number of generated circuits that are needed to
detect these 13 bugs is less than 850.}

We observe that the bug-finding time varies greatly across different
pipelines. This is expected since pipelines differ structurally, in
the way they are executed as well as in the programming language in
which they are implemented. The median time for a single pipeline run
is 0.5s for \circom, 0.1s for \corset, 3s for \gnark, and 6s for
\noir. Thus, even for the same number of generated circuits, \gnark
and \noir are expected to have higher bug-finding times.
For instance, \gnark is essentially a library, and a pipeline run is a
sequence of API calls that need to be compiled before the circuit
compilation; similarly, a \noir pipeline run performs additional
analysis of circuits and executes a virtual machine.
Of course, there are also differences across pipelines in how much
time is spent in each stage. This affects how often the faster and
slower stages are executed, which in turn impacts the efficiency of
\tool.

It is also worth noting that \emph{all bugs are detected using the
default configuration of \tool, which generates small and relatively
simple circuits}. Consequently, the percentage of SAT circuit inputs
(shown in the fourth column of \tabref{tab:efficiency}) is always
greater than 52\% despite the fact that inputs are generated using
blackbox fuzzing. Such a high percentage is important for the fuzzer's
effectiveness since UNSAT circuit inputs typically cannot exercise
later stages of the pipeline, like proof generation and
verification. A significantly smaller percentage would, therefore,
introduce unwanted bias towards the earlier pipeline stages.

\paragraph{RQ4: Design choices and settings.}

\begin{figure}[t]
  \includegraphics[scale=0.5]{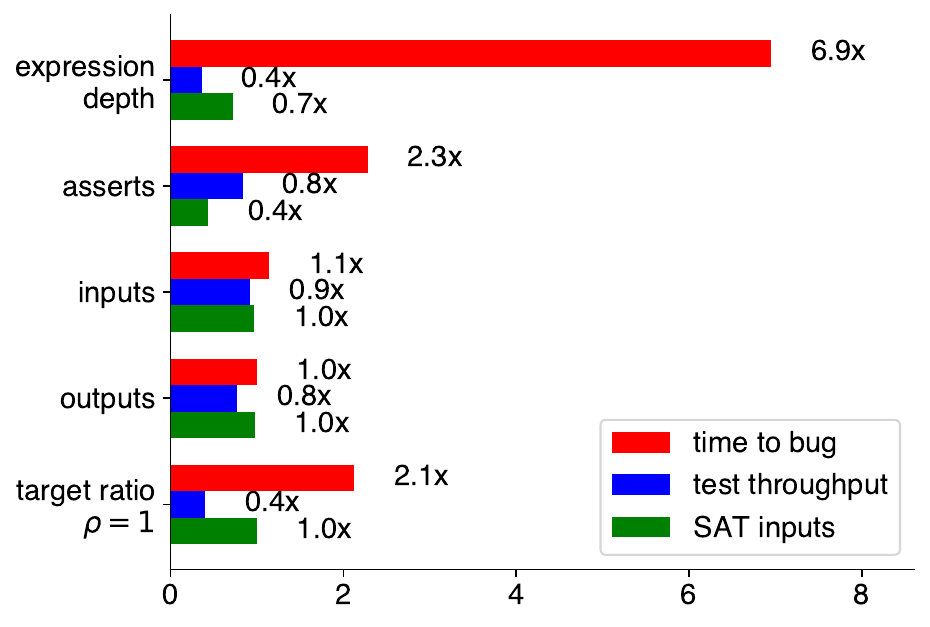}
  \caption{Comparison of the \tool default configuration with five
    variants. The bars show the relative change with respect to three
    metrics: (1)~bug-finding time, (2)~test throughput, and (3)~the
    percentage of SAT inputs.}
  \label{fig:compare-configs-geo-means}
  \Description{}
\end{figure}

In this research question, we evaluate the effectiveness of several
design choices and settings in \tool. Specifically, we consider five
variants of the default configuration, each of which modifies a single
setting: (1)~maximum number of \emph{inputs} (changed from 2 to 8),
(2)~maximum number of \emph{outputs} (changed from 2 to 8),
(3)~maximum number of \emph{assertions} (changed from 2 to 8),
(4)~maximum \emph{expression depth} (changed from 4 to 16), and
(5)~target ratio $\rho$ (changed from 0.5 to 1.0 to disable the
optimization). For this evaluation, we again run 10 independent
fuzzing campaigns for each of the five configuration variants (with
the same random seeds as for the default configuration).

We focus our comparison with the default configuration (see
\tabref{tab:efficiency}) on the following metrics: (1)~bug-finding
time, (2)~test throughput (i.e., number of generated circuits per
second), and (3)~percentage of SAT inputs. For each metric, we first
compute the median ratio of the variant over the default configuration
for each of the fixed bugs (excluding timeouts), and then calculate
the geometric mean across all bugs. The results are summarized in
\figref{fig:compare-configs-geo-means}.

Unsurprisingly, configurations that generate larger and more
constrained circuits---with \emph{increased expression depth and number of
assertions}---take 6.9x and 2.3x longer to find the given bugs, and
there is a significant reduction in test throughput (0.4x and 0.8x) in
comparison to the default configuration. Moreover, the variant with
increased expression depth timed out for 7 out of 10 seeds for
bug~\bug{12}, and for 2 seeds for bug~\bug{14}. The variant with an
increased number of assertions timed out for 1 seed for
bug~\bug{14}. The only outliers in terms of bug-finding time are
\noir's bugs~\bug{13} and~\bug{15}. Both require several nested
expressions to be detected. As a result, they are found faster when
increasing the expression depth.

With the default configuration, $\sim$59\% of all circuits with
concrete inputs were SAT (see \tabref{tab:efficiency}). Larger and
more constrained circuits predictably result in fewer SAT inputs: when
increasing the expression depth, only $\sim$42\% of all generated
inputs were SAT, and when increasing the number of assertions, only
$\sim$26\% of all inputs satisfied the corresponding constraint
systems.

\emph{Increasing the number of circuit inputs} does not have a drastic effect
on any of our metrics. This is expected since our circuit-generation
step does not force all inputs to be used. Therefore, simply adding
more signals does not necessarily make the generated circuits more
complex.

\emph{Increasing the number of circuit outputs} does not have a significant
effect on the bug-finding time. However, it slightly reduces the test
throughput. Upon closer inspection, we observed that the bug-finding
time increased by 2.9x for \corset, but decreased for other pipelines
(0.5x for \circom, 0.9x for \gnark, and 0.7x for \noir). Recall that
\corset does not directly support outputs, and our translation simply
introduces additional temporary variables. As a result, increasing the
number of outputs increases the complexity of \corset circuits. For
other pipelines however, the additional complexity is offset by making
the oracle more effective---more output values are included in each
generated witness, and thus, more data can be compared between the two
witnesses of semantically equivalent circuits.

After considering the four variants that change settings in the
circuit generator, let us now consider the final variant that
\emph{increases the target ratio $\rho$ from 0.5 to 1.0}, thereby disabling
the optimization that frequently skips the later, slower stages of a
pipeline. In other words, this variant always runs the entire
pipeline, whereas the default configuration (with the optimization)
tries to balance the time that is spent in the earlier and later
pipeline stages.

With the optimization, we would expect that only a small portion of
all generated circuits execute the entire pipeline. For three
pipelines, this expectation is also confirmed experimentally. For
instance, \gnark's prover is significantly slower than the rest of the
pipeline, and only 0.7\% of all generated circuits execute the entire
pipeline. \circom and \noir's provers are also slow, and only 3\% (for
\circom) and 10\% (for \noir) of all circuits execute the entire
pipeline. In contrast, for \corset, the full pipeline is executed for
35\% of all circuits. This higher percentage results from the fact
that we include the optional check stage (checking the validity of the
constraint system that is generated from a given circuit) in the
earlier pipeline stages, thereby making them more computationally
expensive. To compensate, the later stages run more often for \corset
than for other pipelines.

When disabling the optimization, we observe higher bug-finding
times. The number of timeouts also increased without the optimization:
for bug~\bug{12} (6 out of 10 seeds timed out), bug~\bug{13} (1 out of
10), and bug~\bug{14} (2 out of 10). This is not surprising since the
test throughput dropped to 0.4x in comparison to the default
configuration.

As these results show, the default configuration provides a good
trade-off for effectively finding bugs across these different
pipelines.

\subsection{Threats to Validity}
\label{sect:threats}

Our experimental results depend on the ZK pipelines under test, their
settings (e.g., the used curve or optimization level), and the
settings of \tool. Moreover, fuzzing per se is a random process, and
\tool randomly generates circuits, transformations, etc. To address
these potential threats, we selected four ZK pipelines that differ in
their circuit programming language, their architecture, and backend
components. In addition, we randomized the settings of the pipelines
(via metamorphic transformations) and systematically varied settings
of \tool during our evaluation. To mitigate effects of randomness on
fuzzing, we ran 10 independent fuzzing campaigns for each evaluated
\tool configuration.


\section{Related Work}
\label{sect:related}

We present the first systematic fuzzing technique for
circuit-processing pipelines. It uses metamorphic test
oracles~\cite{ChenCheung1998,BarrHarman2015,SeguraFraser2016} to find
logic bugs. To the best of our knowledge, there is no existing work on
fuzzing circuit-processing pipelines, but the problem
itself has (independently) been described in the literature as an open
problem~\cite{ChaliasosErnstberger2024}.

There is recent work on finding bugs in a \emph{circuit
itself}~\cite{WenStephens2024}. In contrast, our approach focuses on
detecting bugs in a \emph{circuit-processing pipeline}. Both types of
bugs could have catastrophic consequences, but a bug in a
circuit-processing pipeline may affect many, or even all, deployed
circuits.

The most closely related areas to our work are fuzzing for
compilers~\cite{ChenPatra2020} and program
analyzers~\cite{MidtgaardMoeller2017,KapusCadar2017,BugariuWuestholz2018,KlingerChristakis2019,ZhangSu2019,BugariuMueller2020,MansurChristakis2020,WintererZhang2020-Operators,WintererZhang2020-Fusion,TanejaLiu2020,ParkWinterer2021,IrfanPorncharoenwase2022,EvenMendozaSharma2023,MansurWuestholz2023,ZhangPei2023,ZhangPei2024,HeDi2024,KaindlstorferIsychev2024,FleischmannKaindlstorfer2024},
such as software model checkers~\cite{BiereCimatti1999,McMillan2018}
and abstract interpreters~\cite{CousotCousot1977}.
After all, the
first stage in circuit-processing pipelines typically invokes a
compiler for the ZK language. Similarly, most program analyzers have a
compiler frontend that parses the input programs and often translates
them into an intermediate language used for the analysis. The
translation could generate a control-flow graph (as in many dataflow
analyzers~\cite{Kildall1973} and abstract interpreters), a program in
an intermediate verification
language~\cite{BarnettChang2005,FilliatrePaskevich2013,MuellerSchwerhoff2016}
(as in many deductive verifiers, such as Dafny~\cite{Leino2010} and
Spec\#~\cite{BarnettFahndrich2011}), or a GOTO program (as in
CBMC~\cite{BiereCimatti1999} and several other software model
checkers).

On the other hand, regular compilers typically translate from a
high-level language (such as C) to a more low-level language (such as
assembly or LLVM bitcode). In contrast, ZK pipelines translate to a
constraint system, and there are several later stages that make heavy
use of cryptographic primitives for generating and verifying
proofs. ZK pipelines are, therefore, highly complex and may contain
even more subtle and hard-to-detect bugs than regular compilers.

Generally, most existing work on fuzzing for compilers and program
analyzers uses one or more of the following three types of oracles:
(1)~specification-based oracles~\cite{BarrHarman2015} (by comparing
the actual behavior to a formal specification of the expected
behavior), (2)~differential oracles~\cite{McKeeman1998,BarrHarman2015}
(by comparing the behavior of two or more implementations), and
(3)~metamorphic oracles. In principle, all of these types of oracles
could be used for ZK pipelines. In this work, we have mainly focused
on metamorphic oracles and have used specification-based oracles to
express correctness properties of the pipelines. In the future, we
plan to extend our fuzzer to also perform differential testing by
comparing multiple pipelines. We also plan to incorporate more metamorphic transformations
for settings, which can also be viewed as a limited form of differential testing; for
instance, by enabling different proof systems via flags.


\section{Conclusion}
\label{sect:conclusion}

We have presented \tool, the first fuzzer for detecting logic bugs in
circuit-processing pipelines.
It introduces \il, an intermediate language for circuit generation,
and rewrite rules over this language for metamorphic circuit
transformations. \tool translates the generated and transformed
circuits into the ZK language of the pipeline under test and generates
inputs for them using blackbox fuzzing. Bugs are detected by executing
the pipeline under test on the circuits and checking for violations of
the metamorphic oracles and other non-metamorphic, correctness
properties.
We used \tool to test four diverse ZK pipelines and
detected critical bugs in all of them.

Despite the bug-finding effectiveness of \tool, there are still
several (optional) components of ZK pipelines that are not being
tested. For example, as an alternative to the existing verifier,
\gnark allows generating a Solidity contract, which, when compiled and
deployed on the blockchain, could also be invoked to verify a
generated proof. In this example, even bugs in the generated Solidity
contract, the Solidity compiler, or the Ethereum virtual machine could
compromise the correctness of the extended ZK pipeline. As a next
step, we plan to explore how to test such components.



\begin{acks}
We are grateful to the ZK-pipeline developers for their valuable help.
This work was supported by the Vienna Science and Technology Fund
(WWTF) and the City of Vienna [Grant ID: 10.47379/ICT22007] as well as
Maria Christakis’ ERC Starting grant 101076510.
\end{acks}

\bibliographystyle{ACM-Reference-Format}
\bibliography{bibliography}

\appendix

\end{document}